\documentclass[11pt]{amsart}

\usepackage[left=3cm, right=2cm, top=3cm, bottom=2cm]{geometry}                
\usepackage[dvipsnames]{xcolor}
\usepackage{graphicx}
\usepackage{amssymb}
\usepackage{epstopdf}
\usepackage{caption}
\usepackage{float}

\usepackage{listings}
\usepackage{amsmath}
\usepackage{verbatim}
\usepackage{caption}

\usepackage{afterpage}

\newcommand\blankpage{%
    \null
    \thispagestyle{empty}%
    \newpage}

\title{AWAKE  \\ 
on the path to particle physics applications }
\author{AWAKE Management Team \\
A.~Caldwell$^1$, E.~Gschwendtner$^2$, K.~Lotov$^{3,4}$, P.~Muggli$^{1,2}$, M.~Wing$^5$} 

\date{17-December-2018}

\begin{document}
\maketitle
\vskip -2.cm

\begin{abstract}
    Proton-driven plasma wakefield acceleration allows the transfer of energy from a proton bunch to a trailing bunch of particles, the `witness' particles, via plasma electrons.  The AWAKE experiment at CERN is pursuing a demonstration of this scheme using bunches of protons from the CERN SPS.  Assuming continued success of the AWAKE program, high energy electron or muon beams will become available, opening up an extensive array of future particle physics projects from beam dump searches for new weakly interacting particles such as Dark Photons, to fixed target physics programs, to energy frontier electron-proton, electron-ion, electron-positron and muon colliders.
    The time is right for the particle physics community to offer strong support to the pursuit of this new technology as it will open up new avenues for high energy particle physics. 
\end{abstract}

\vskip 10.cm
\noindent\address[1]{Max Planck Institute for Physics, Munich, Germany} \\
\address[2]{CERN, Geneva, Switzerland} \\
\address[3]{Budker Institute of Nuclear Physics SB RAS, 630090, Novosibirsk, Russia} \\
\address[4]{Novosibirsk State University, 630090, Novosibirsk, Russia} \\
\address[5]{University College London, London, UK }\\
Contact: Allen Caldwell (caldwell@mpp.mpg.de)
\blankpage

\section{Introduction}
Now that the Higgs has been discovered, the question `what are the most promising directions of research in high-energy particle physics?' is being posed with ever increasing urgency.  We all know the list of `known questions', such as `what is the nature of dark matter and dark energy, what is the cause of the matter-antimatter imbalance, $\cdots$?' and there are surely many `unknown questions' - the questions that we should be posing to discover new physics.  Particle accelerators have played a major role in the development of our understanding of nature in the past.  They have the potential to reveal more of nature's secrets in the future also.

An oft-stated goal of novel accelerator technology is to provide a means to build a high energy electron-positron linear collider capable of reaching high luminosity at TeV and beyond energy scales at a much lower cost than could be achieved with existing radio frequency based accelerating structures.  While achieving this would indeed provide an exciting outlook for particle physics, we do not consider this as the only goal of developing novel accelerator technology.  Rather, the technology development should be seen as a generic way of providing research opportunities.  As has often been the case in the history of physics, new technologies bring discoveries - often not along the lines for which they were initially developed.  Novel accelerators based on plasma wakefields, dielectric structures or direct laser acceleration can bring scientific opportunities that reach far beyond colliders, and should therefore be pursued with the highest priority as enablers of discoveries. 


 An important and promising future technology is proton-driven plasma wakefield acceleration (PDPWA).  As we demonstrate in this document, it has the potential to greatly extend the use of existing proton accelerators for myriad particle physics applications.  We first review the AWAKE scheme to realize PDPWA and give the main results from Run 1 of the experiment, which has just ended.  We then outline the proposed Run 2 of AWAKE, which will bring the developments to a point where particle physics projects can be realized.  A first round of studies that have been performed on particle physics applications of the AWAKE-like scheme is then summarized.   These studies indicate the great potential that is present already in the midterm (5-10 years) and that could be realized with strong support from the accelerator, plasma and particle physics communities.
 \blankpage

\section{Proton-driven plasma wakefield acceleration}

The potential of plasma as the medium for high gradient acceleration has been demonstrated with short and intense 
laser pulse drivers yielding electron bunches of up to 4.25\,GeV energy gain in cm-long channels~\cite{leemans,wang} 
that corresponds to about 100\,GV/m average accelerating fields.  High gradients have also been demonstrated with 
a short, high charge electron bunch driver with an energy gain of 42\,GeV in 85\,cm, corresponding to 
52\,GV/m~\cite{slac}.

However, in both of these pioneering experiments the energy gain was limited by the energy carried by the driver 
and the propagation length of the driver in the plasma ($<1$\,m). The laser pulse and electron bunch driver 
schemes therefore require staging~\cite{staging,leemans-stage}; i.e., the stacking of many $10-25$\,GeV acceleration,  
or plasma, stages to reach the $\sim 1$\,TeV energy per particle or equivalently $\sim 2$\,kJ of energy in 
$\sim 2 \times 10^{10}$ electrons and positrons. The AWAKE scheme avoids these propagation-length and 
energy limitations by using a proton bunch to drive the wakefields. 

Bunches with $3 \times 10^{11}$\,protons and 19\,kJ of energy (the CERN SPS 400\,GeV beam), and with 
$1.7 \times 10^{11}$\,protons and 180\,kJ of 
energy (the CERN LHC 6.5\,TeV beam) are produced routinely today. Because of their high energy and mass, proton 
bunches can drive wakefields over much longer plasma lengths than other drivers. They can take a witness bunch to 
the energy frontier in a single plasma stage, as was demonstrated in simulations~\cite{ref:caldwell09}. This 
proton-driven scheme therefore greatly simplifies and shortens the accelerator. In addition, because there is no gap 
between the accelerator stages, this scheme avoids gradient dilution.

\subsection{Self-modulation of particle beams in plasma}
The major challenge in realizing proton-driven plasma wakefield acceleration today is the length of the existing proton bunches. The maximum field of a plasma wake 
scales as

\begin{equation}
E_{\rm max} \approx \sqrt{\frac{n_e [{\rm cm}^{-3}]}{10^{14}}} \, {\rm GV/m},
\end{equation}
where $n_e$ is the plasma electron density. Consequently, plasma densities of at least $n_e \approx 10^{14}\,$cm$^{-3}$ 
are required to reach accelerating gradients of GV/m and above. The corresponding plasma wavelength, calculated from the plasma oscillation frequency and assuming a relativistic driver, is

\begin{equation}
\lambda_p \approx \sqrt{\frac{10^{15}}{n_e [{\rm cm}^{-3}]}} \, {\rm mm}.
\end{equation}
At these densities, the plasma wavelength is of the order of a millimeter. On the other hand, proton bunches available 
today are much longer, $\sigma_z = 3-12$\,cm.

\begin{figure}[htp]
\includegraphics[width=0.8\textwidth]{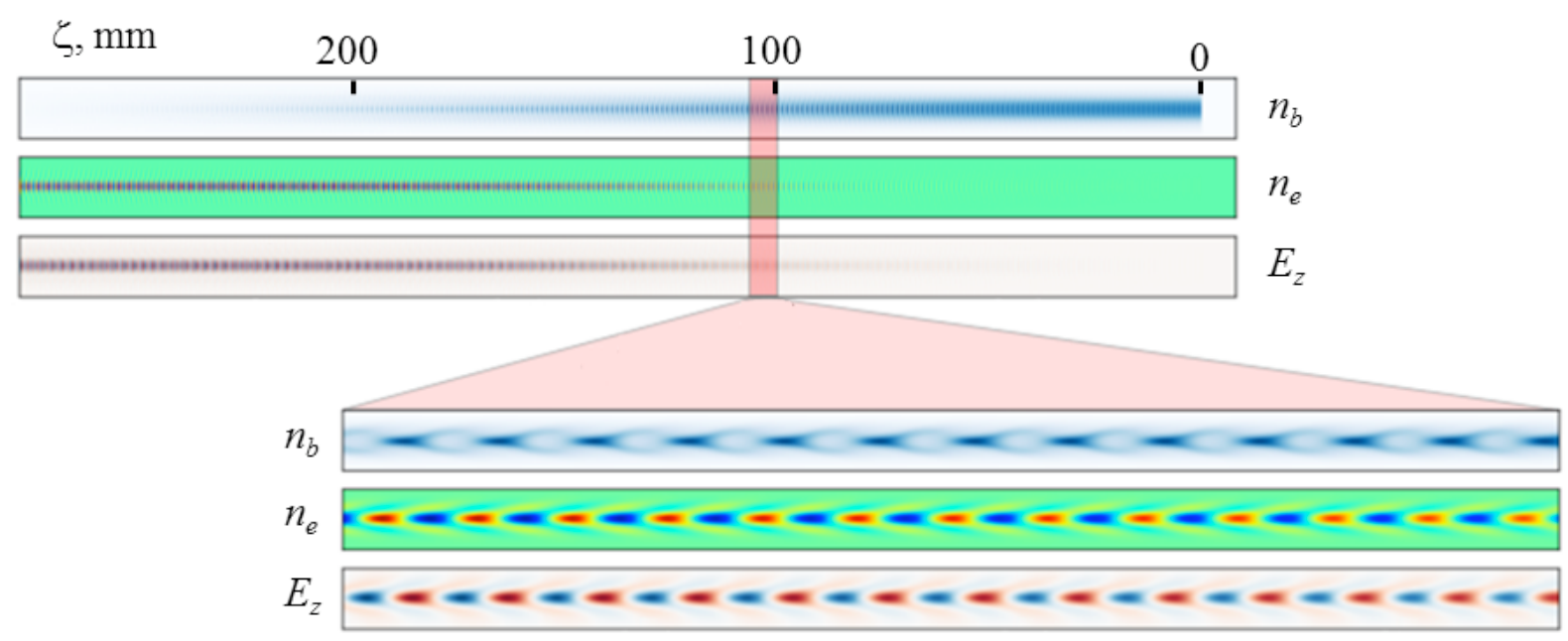}
\caption{Proton beam density $n_b$, plasma electron density $n_e$, and the longitudinal component of the wakefield 
$E_z$ after 4\,m of propagation in plasma. The coordinate $\zeta = ct - z$ is counted from the bunch head. The region 
10\,cm behind the bunch head is zoomed in.}
\label{fig:smi}
\end{figure}

A mechanism has been discovered that automatically splits the proton bunch propagating in plasma 
into a number of micro-bunches  via self-modulation~\cite{smi-lotov,smi-pukhov}. The process starts 
from a seeding wave whose transverse field acts on the beam and modulates its radius.  Its amplitude grows exponentially from head to tail of the bunch and along the 
propagation distance.  The physics of the process is now well understood, and theoretical predictions agree well 
with results of simulations~\cite{smi-sat1,caldwell-lotov}.  At saturation, the initially long 
and smooth beam is split into a train of micro-bunches that resonantly excite a strong plasma wave.  This plasma wave 
is inherently weakly nonlinear~\cite{PoP20-083119}, so its excitation is almost independent of the charge 
sign of the drive beam.  An example of a self-modulated bunch as observed in 3D simulations using 
the particle-in-cell code VLPL~\cite{vlpl1,vlpl2} is shown in Figure~\ref{fig:smi}.

\subsection{Results from AWAKE Run 1}

\begin{figure}[h]
\includegraphics[width=\textwidth]{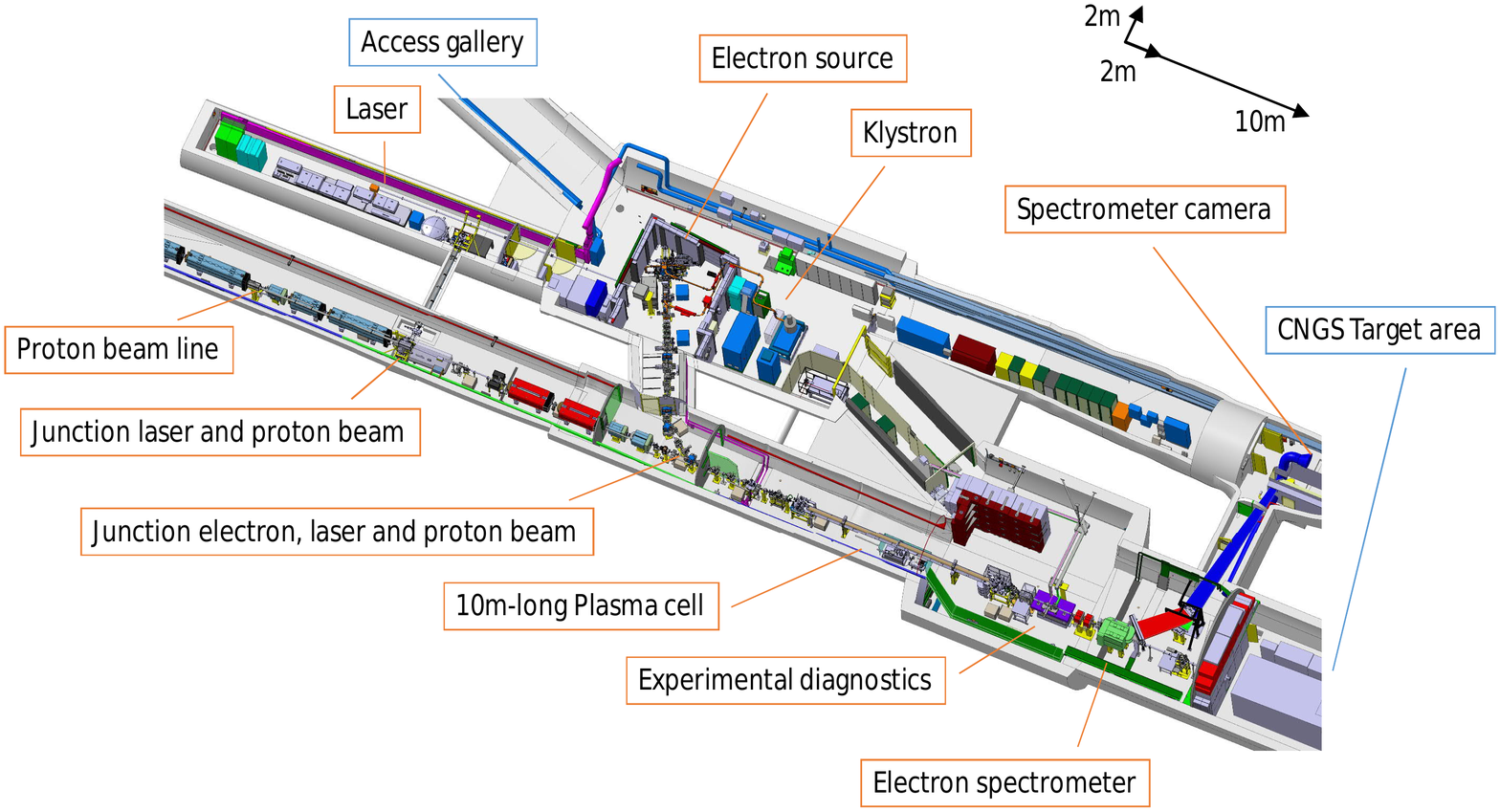}
\caption{Layout of the AWAKE experiment.}
\label{fig:design}
\end{figure}

The layout of the AWAKE experiment is shown in 
Figure~\ref{fig:design}. The laser ionizes a Rubidium vapor 
in the first plasma section and seeds the self-modulation process~\cite{oz-muggli}.  The electrons are injected in the wakefields and their energy is measured with 
an electron spectrometer.

First protons to the experiment were delivered at the end of 2016 and AWAKE has now completed a two year experimental program.

\subsubsection{Self-modulation process and seeded self-modulation studies}

\begin{figure}[ht]
\centering
\includegraphics[scale=0.5]{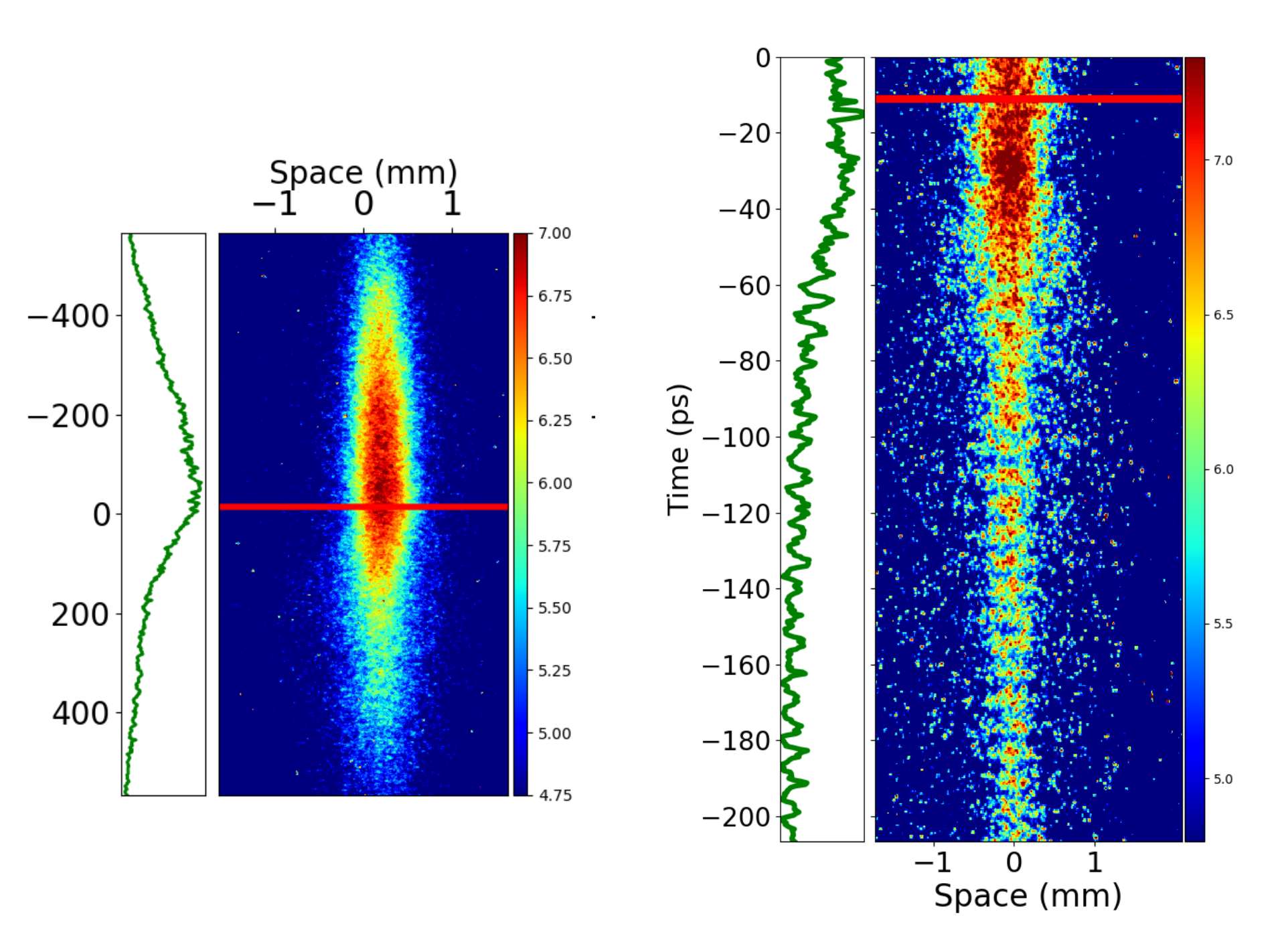}
\caption{Left: streak camera image of the proton bunch at the 1\,ns time scale showing the effect of
 the plasma starting at the laser pulse time indicated by the red line (0\,ps, front of the bunch at
 the top). %
The current density profile of the bunch is shown by the green line. %
Right: similar image, zoomed in (210\,ps time scale) clearly showing the microbunches resulting from the SSM development after the ionising laser pulse and start of the plasma.}
\label{fig:ssm} 
\end{figure}

Throughout 2017, AWAKE acquired data on the physics of proton self-modulation by the plasma wakefield under a variety of experimental 
conditions. The results~\cite{SSM} demonstrate for the first time the observation of the seeded self-modulation (SSM) of 
a long charged particle bunch in a plasma and confirm many of the theoretical predictions about the SSM.  The SSM manifests itself by the formation of a microbunch train with periodicity equal to the electron plasma frequency. %
This is demonstrated from short time scale (73 and 200\,ps) streak camera images 
showing microbunched structure starting at the location of the plasma creation along the bunch (see 
Fig.~\ref{fig:ssm}). %
The periodic structure extends over more than one rms length of the proton bunch. %

\subsubsection{Electron injection and acceleration studies}

\begin{figure}[ht]
\centering
\includegraphics[scale=0.4]{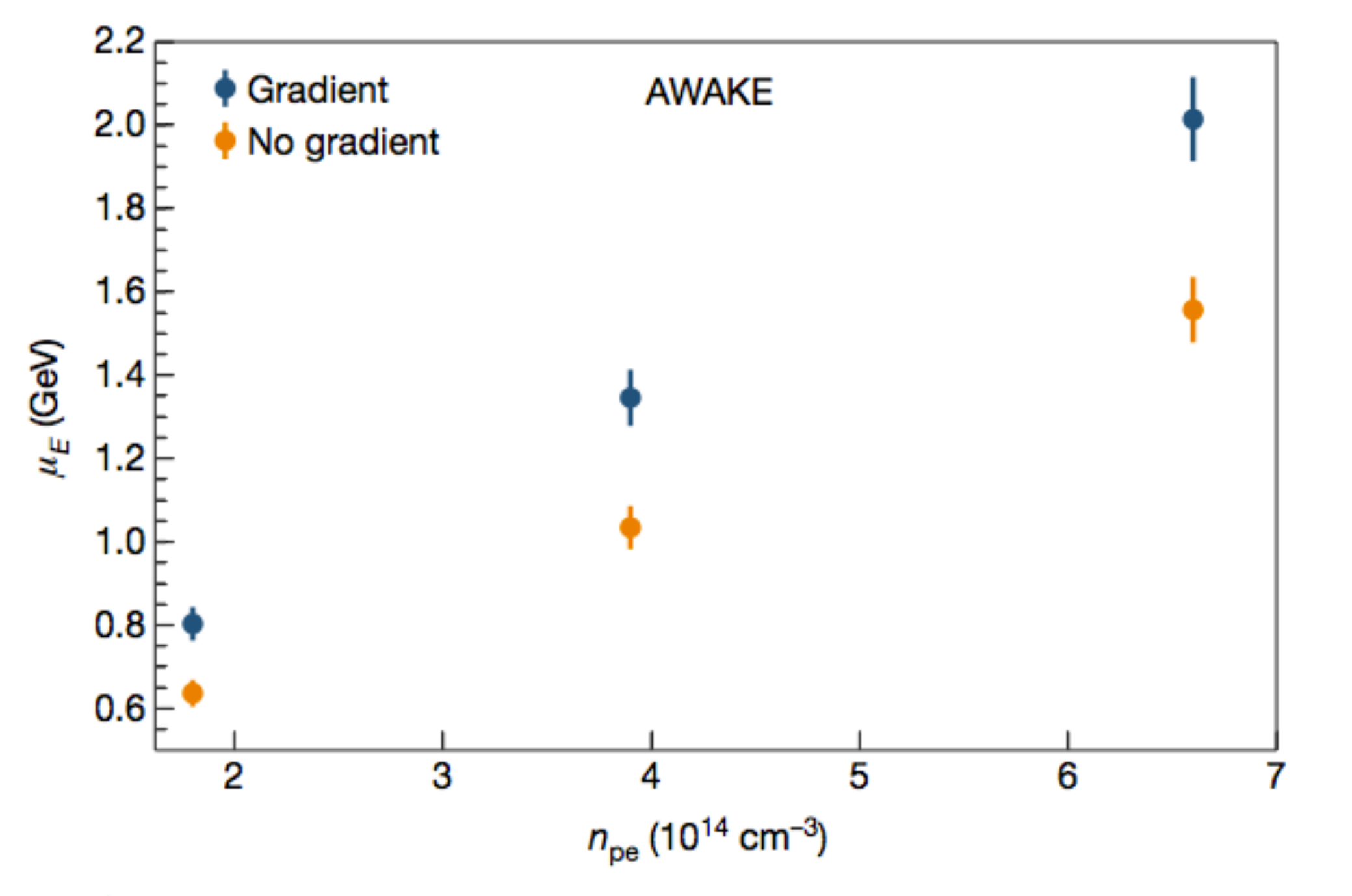}
\caption{Energy of externally injected electrons with $\sim$19\,MeV after the 10\,m plasma as a function of plasma density without (blue circles) and with (orange circles) a positive plasma density gradient (of $2.2$~\% for the point at $2.0$~GeV). }
\label{fig:accel} 
\end{figure}

Starting in 2018, electron injection experiments began in earnest and acceleration of electrons up to 2 GeV was observed~ \cite{ref:electron}. The measurements show that, as expected, the energy gain increases with plasma density (see Fig.~\ref{fig:accel}). %
Acceleration in a plasma with a positive density gradient of a few percents over 10\,m show larger energy gains. %
These results are in agreement with the expectations from simulations.

These two key results, seeded self-modulation of the proton bunch and acceleration of externally injected electrons, form the basis for the next experiments aiming to demonstrate the production of a high energy, high quality, electron bunch.

\newpage

\section{AWAKE Run 2: on the path to particle physics applications}

The goals for Run 2 are to bring the R\&D development of PDPWA to a point where particle physics applications can be proposed and realized.  For this, in  Run 2 AWAKE will: 
\begin{enumerate}
\item 
demonstrate emittance control of an electron bunch during its merging into the modulated proton bunch and during its acceleration in the plasma wakefield produced by the proton micro-bunches with stabilized emittance at the  $10$~mm-mrad level;
\item
demonstrate scalable acceleration with gradients in the range $0.5-1$~GV/m for an accelerated bunch charge at least $100$~pC.
\end{enumerate}

\begin{figure}[th]
\centerline{\includegraphics[width=0.7\paperwidth]{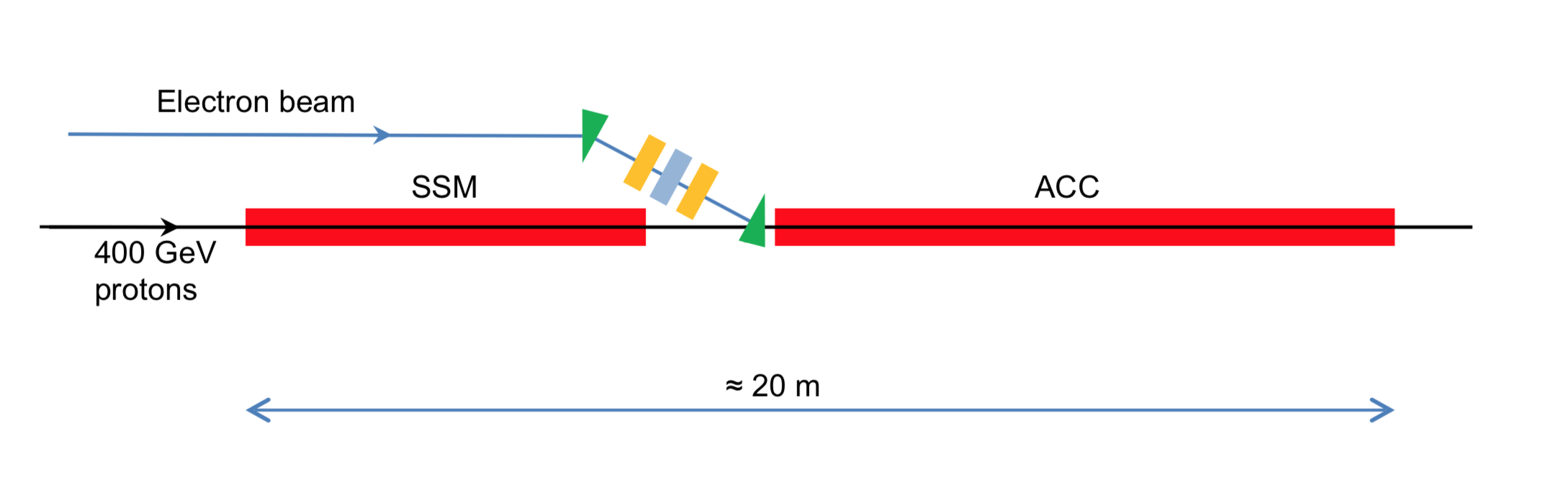}}
\vskip -1cm
\emph{\caption{\protect\label{fig:layout-1} The conceptual
layout for the plasma sources and electron injection in AWAKE Run 2. The first plasma source is used to modulate the proton bunch (SSM section). The electron bunch is injected after the proton bunch has modulated, and accelerated in the downstream plasma (ACC section). }
}
\end{figure}

In Run 1, electron injection was performed near the entrance to the single plasma source used in the experiment (see Fig.±\ref{fig:design}), and before significant proton bunch modulation had taken place. The incoming electron beam energy was limited to $20$~MeV and the bunch length was of order of the plasma wavelength. The goals for Run 2 require that the electron bunch be injected into the modulated proton bunch {\bf after} the SSM process has saturated and been stabilized (see Fig.±\ref{fig:layout-1}).  They also require that the electron bunch that is injected has {\bf sufficient charge density} that beam loading can occur.  
It is also intended to implement a step in the plasma density seen by the proton bunch during the SSM process: this will freeze the modulation process and allow for much higher final electron energies over long acceleration distances~\cite{caldwell-lotov}.

Simulation studies show that a beam of
a few 100\,pC charge, with $40-60$\,$\mu$m bunch length and an emittance of
order 2\,$\mu$m, matched to the plasma (beta function of order few
mm), can be accelerated with small emittance growth if injected correctly
into the plasma~\cite{SPSC_BEAMLOAD}. A new, higher energy electron
injector and transport line will be required to generate an electron
beam with these parameters within an acceptable distance. 

It is of great importance for future applications
to develop plasma source technologies that are as cost effective
as possible if plasma accelerators of 100s or 1000s of meters in length
are to be built. A scalable technology should as well have negligible
or no vacuum gap between the plasma source modules to avoid staging
issues~\cite{SPSC_ISTAGE}.  A collaboration to develop scalable helicon
plasma sources as well as discharge plasma sources at CERN has been
set up~\cite{SPSC_HELICON}.  This plasma source development effort will benefit the whole advanced accelerator community since all plasma-based colliders will require meter-scale, high quality plasma sources. 

Run 2 is foreseen to take place between LS2 and LS3 of the LHC; i.e., from 2021-2024.  Given the continued successful development of the PDPWA scheme pursued by AWAKE, it will be possible to propose and realize particle physics projects based on this approach.  Potential applications should therefore be developed in parallel to the accelerator R\&D carried out by AWAKE to realize these projects in a timely way.

\section{Particle Physics Applications}

Using SPS proton bunches as drivers, electron energies in the range $50-100$~GeV are expected to be possible, whereas TeV-scale electrons should be possible using LHC proton bunches as drivers.  These values are considered in the physics applications that follow\cite{MatthewESPP}.  The AWAKE scheme is also promising in conjunction with future accelerator concepts such as the FCC, the muon collider, and a gamma factory based on partially stripped ions.  We discuss these briefly at the end of this section.  The feasibility of for-purpose designed proton accelerators, i.e. with short ($<1$~mm)  bunches, has not yet been investigated, but should be part of a long-term strategy.


\subsection{PDPWA for beam dump experiments}

One generic class of experiments that can be carried out with high energy particle beams are beam dump experiments, where new, weakly interacting particles are searched for behind an absorber used to filter out the bulk of known particles.  High incident particle fluxes allow for increased sensitivity in the coupling strength to known particles, while increased beam energy increases the reach in mass of the new particles.  The requirements on the emittance of the accelerated electrons are very mild in this case, and not foreseen to be difficult to reach.



\begin{figure}[h]
\begin{center}
\includegraphics[width=0.6\textwidth]{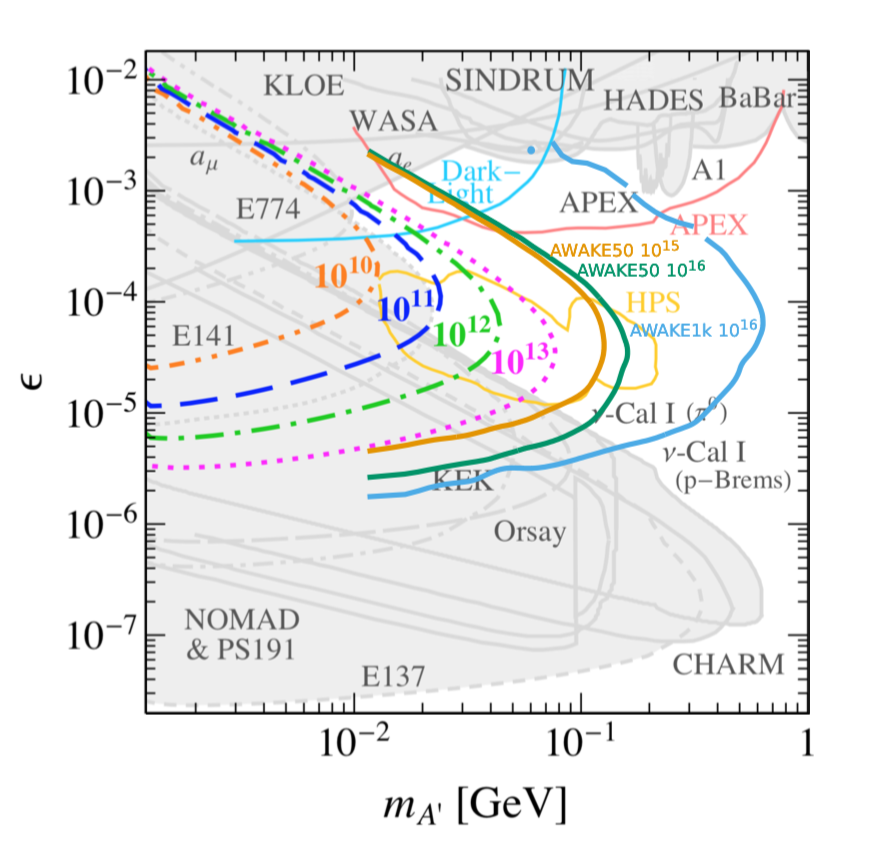}
\caption{The existing limits on dark photons are shown in gray shading, together with exclusion limits assuming $50$~GeV electron bunches (colored lines) for different assumptions on the numbers of electrons on target.  The blue curve indicating sensitivity to higher masses is based on $1$~TeV electron bunches. The limits are shown in the plane of dark photon mass, $m_A'$, and Standard-Model mediator coupling $\epsilon$.  From ~\cite{ref:Anthony}. }
\label{fig:Dark}
\end{center}
\end{figure}

As an example, we consider here the search for Dark Photons.  The nature of Dark Matter is a persistent mystery, and the possibilities for the particle content of the 'Dark Sector' are many.  One such possibility is the existence of 'Dark Photons', which could further decay into dark matter particles or Standard Model particles~\cite{ref:DS}.  A study, making use of SPS proton bunches to drive electron bunches~\cite{MatthewESPP}, assumed one 12-week data taking period resulting in $10^{16}$,~$50$~GeV electrons on target.  The beam line and integration of the fixed target experiment is presented in another document submitted to the ESPP update \cite{EddaESPP}. The resulting 90~\% confidence limit contour is shown in Fig.~\ref{fig:Dark} in the parameter space of the mass of the dark photon, $m_{A'}$, and the dimensionless coupling to Standard Model particles, $\epsilon$ in decays to $e^+e^-$.  The experimental conditions were modeled on an experiment such as the  NA64 experiment~\cite{ref:NA64}. A significant extension of the kinematic coverage is found.  The availability of much higher electron energies using the LHC bunch as driver of the wakefields further  greatly extends the range to higher $m_{A'}$ masses as shown by the blue curve in Fig.~\ref{fig:Dark}. This represents one example of a search for new physics in beam dump experiments that would benefit greatly from the availability of high energy electron beams. 

\subsection{PDPWA for fixed target experiments}

The highest energy lepton-hadron fixed target experiments have been performed at FNAL and CERN.  The lepton beams are produced by scattering high energy protons on a target and extracting the leptons in a large momentum band.  As an example, the M2 beam at CERN uses 450 GeV protons from the SPS, with a muon production efficiency of approximately $10^{-5}$ per proton on target.  The momentum range of the muons spans from $60-190$~GeV/c.  Taking $E_{\mu}=160$~GeV results in a center-of-mass energy for $\mu p$ scattering of $17$~GeV.  With the 3~TeV electron beams available from using LHC proton bunches as wakefield drivers, the center-of-mass energy for electron-proton scattering would increase to $75$~GeV, in the middle of the range ($15-140$~GeV) planned for the next generation nuclear physics collider, the EIC.  

As an example, we consider accelerating to $3$~TeV $10^{10}$ electrons in the wake of an LHC proton bunch at an effective repetition rate of 1 Hz for $10^7$ seconds per year in order to measure high $x$ structure functions in the region where perturbative QCD applies.  This region is particularly important in the search for new effects at high center-of-mass energy of the interacting quarks and gluons in proton collisions.  Using a thick liquid hydrogen target of the kind developed for COMPASS~\cite{ref:COMPASStarget} would yield of order $10^7$ events in the range $x>0.6$ and $Q^2>20$~GeV$^2$, per year of running, allowing for high precision structure function measurements in a crucial kinematic range which currently has large uncertainties.

Polarized electron-proton scattering would be possible, as initial studies~\cite{ref:Vieira} indicate that under the right conditions, electron polarization can be maintained during the acceleration process.  Using a polarized target would allow for a spin physics program, albeit at much smaller integrated luminosities than planned for the EIC.  




\subsection{Electron-proton and electron-ion collider physics}

The HERA electron--proton accelerator was the first and so far only lepton--hadron collider worldwide.  No electron-ion collider has been realized to date. %
The LHeC project~\cite{lhec} is a proposed $ep$ collider with significantly higher energy and luminosity than 
HERA with a program to investigate Higgs physics, to search for new physics, and to perform high-precision measurements of parton densities.
With the AWAKE scheme, we can consider a very high energy electron--proton collider, VHEeP, with
$ep$ center-of-mass energy of about 9\,TeV, a factor of six higher than proposed for the LHeC and a factor of 30 higher than HERA~\cite{ref:caldwellwing}. The luminosity will presumably be relatively modest with a target of $10\,{\rm pb}^{-1}$ over the lifetime of the collider.  An initial version of VHEeP, named PEPIC, would use the SPS to drive electrons bunches to about 50~GeV for collision with LHC protons and ions~\cite{EddaESPP}.  PEPIC would be of interest in case the LHeC is not realized.

\subsubsection{QCD physics at VHEeP}
\label{sec:qcd}

Electron--proton collisions at $\sqrt{s} \sim 9$\,TeV give access to a completely new kinematic regime for deep inelastic 
scattering with, in particular, a reach in Bjorken $x$ a factor of about 1\,000 lower than at HERA.  The energy dependence of hadronic cross sections, such as the total 
photon--proton cross section, are not understood and require new experimental results to yield their secrets.   We consider here two example measurements: the measurement of the total photon-proton cross section and the virtual photon-proton cross section as a function of energy.   A huge program of QCD physics would also be available with electron-ion collisions at the ultra-high energies made available by PDPWA.

\begin{figure}[h]
\includegraphics[width=0.7\textwidth]{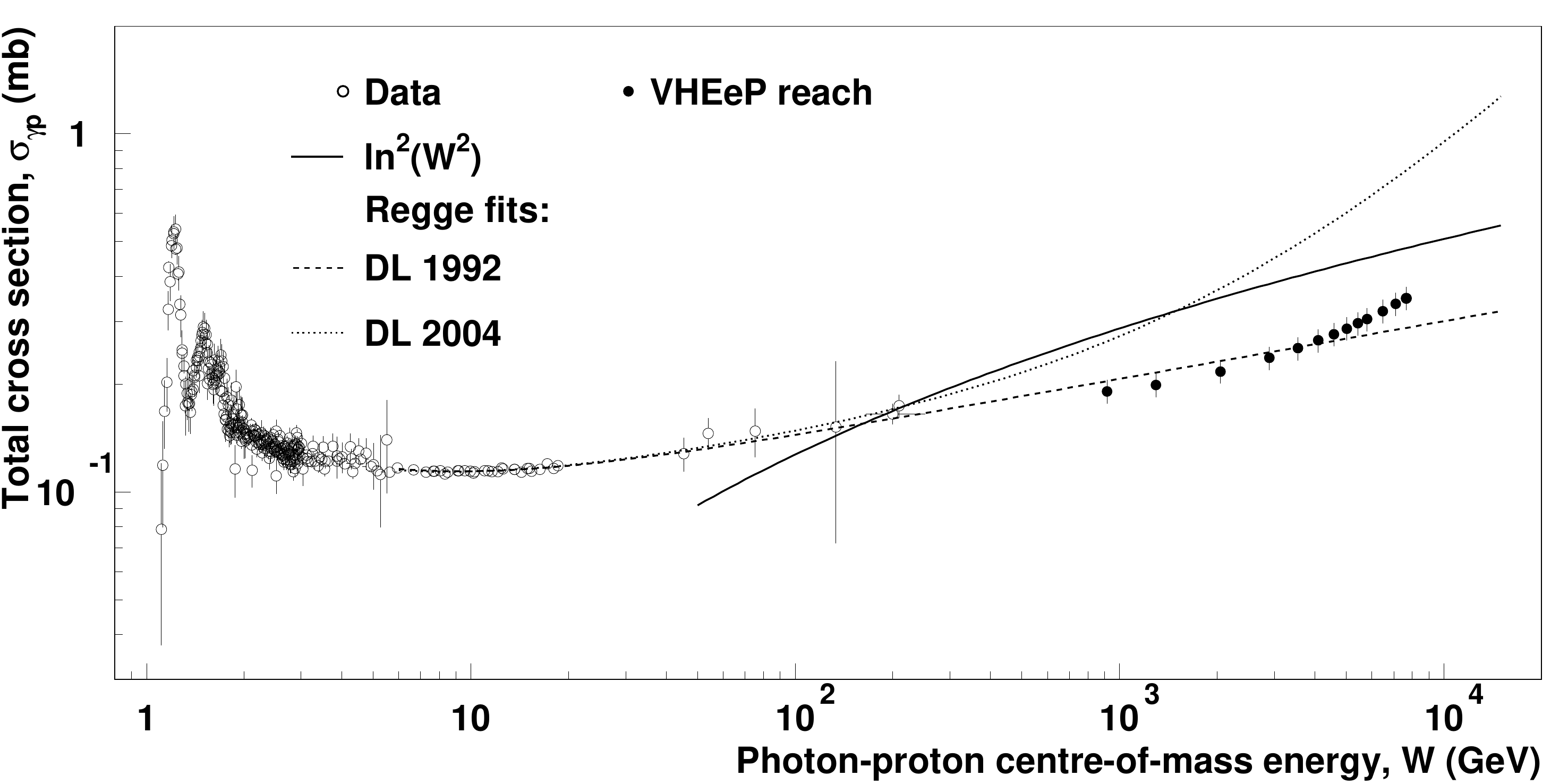}
\caption{Total $\gamma p$ cross section versus photon--proton centre-of-mass energy, $W$, shown for data compared to 
various models.  The data is taken from the PDG, with 
references to the original papers given therein. Data points are also shown for the anticipated $W$ values at which VHEeP data would appear. }
\label{fig:gammap}
\end{figure}

Measurements of the total $\gamma p$ cross section are shown in Fig.~\ref{fig:gammap} compared to phenomenological 
models.  A projection for measurements at the expected values of $W$ at VHEeP is also shown, based on a modest luminosity of about $50$~nb$^{-1}$.  We note that a photon--proton collision of $W = 6$\,TeV, corresponding to photon and proton energies of, respectively, 1.3\,TeV and 7\,TeV, 
is equivalent to a 20\,PeV photon on a fixed target.  This extends significantly 
into the region of ultra high energy cosmic rays and VHEeP data could be used to constrain cosmic-ray air-shower models (for example, see~\cite{pr:d92:114011}).  

\begin{figure}[htp]
\includegraphics[width=0.6\textwidth]{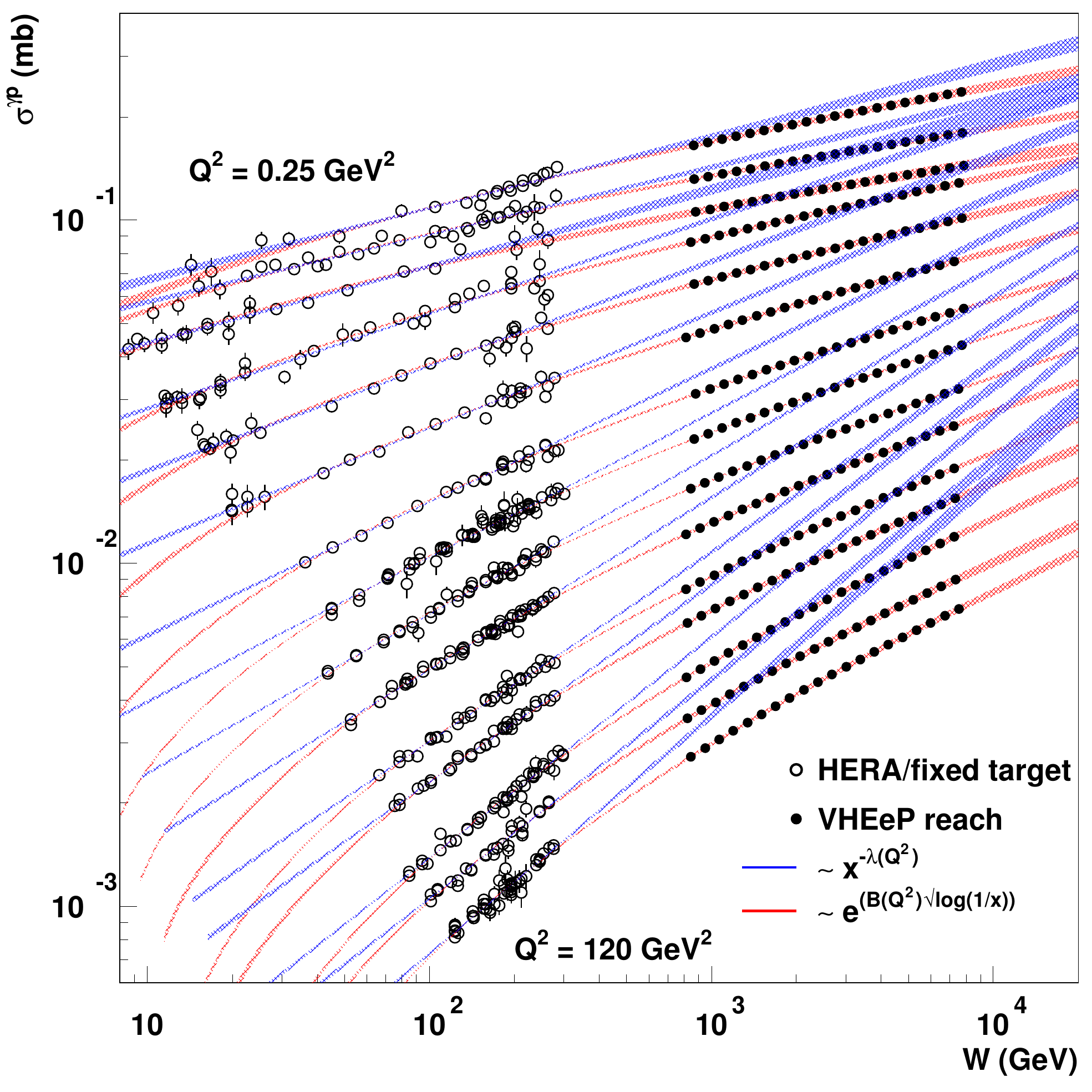}
\caption{Measurements (open points) of $\sigma^{\gamma p}$ versus $W$ for $0.25 < Q^2 < 120$\,GeV$^2$ from HERA and 
fixed-target experiments.  
The blue lines show power law fits to the data, performed separately for each $Q^2$ value.
The red lines show fits of the form inspired by double asymptotic scaling~\cite{ref:BallForte}.}
\label{fig:lowx}
\end{figure}

The energy dependence of 
scattering cross sections for virtual photons on protons is also of fundamental interest, and its study at different virtuality 
is expected to bring insight into the processes leading to the observed universal behaviour of cross sections at high 
energies. Figure~\ref{fig:lowx} shows the results of extrapolation of fits~\cite{ref:caldwell} to the energy dependence of the photon--proton cross section 
for different virtualities, as given in the caption, for two different assumptions on the energy behavior. The reach of VHEeP is shown as projected data points (closed points). The uncertainties are assumed to be of order 1\%, given the increased cross section expected at these energies and assuming similar systematics to those at HERA.

It is clear that VHEeP will yield exciting and unique information on the fundamental underlying physics at the heart of the high energy dependence of hadronic cross sections.

\subsubsection{Physics beyond the Standard Model}
\label{sec:bsm}

The key 
features of VHEeP are the very large center-of-mass energy and the unique combination of electron and proton scattering.  We outline here two areas where VHEeP could yield dramatic breakthroughs - in the search for new particles and in uncovering a possible connection between QCD and Gravity.

New particles with both lepton and quark quantum numbers appear in many extensions of the Standard Model and can be uniquely produced as resonances in a lepton-quark collider. Due to its very large center-of-mass energy, VHEeP sensitivity approaches leptoquark masses of $9$~TeV for electroweak scale coupling strengths, significantly beyond the LHC expectations~\cite{ref:caldwellwing}.  

The connection between gravity and the Standard Model forces remains mysterious. There are intriguing theoretical links between gravity and QCD: the well-known AdS/CFT conjecture links gravity-like theories to QCD-like strong coupling theories~\cite{ref:Maldacena} and the QCD axion is linked to gravitational physics~\cite{ref:Dvali}. In the BFKL approach to small-$x$ physics, the high energy behavior of the DIS cross sections is governed by the Pomeron. The Pomeron, a vacuum-like state of QCD, has properties similar to the graviton in gravitational physics; it shows a non-Wilsonian behavior in that short distances are not necessarily connected to high energy scales. 
In the gravitational interactions, the fundamental objects, high mass black holes, grow in size with energy just as the proton grows in size with energy.  Significantly extending the energy reach of deep inelastic scattering could provide the necessary information to uncover whether these are purely mathematical similarities or whether there is a deep physical link between these forces.

\subsection{Further uses of PDPWA}
Many further investigations of applications of PDPWA are planned.  These include:
\begin{itemize}
\item The use of PDPWA to accelerate bunches of muons to high energies with small muon loss through decay.  In the recently proposed scheme for the front end of a Muon Collider~\cite{ref:LEMMA}, bunches of $6 \cdot 10^9$ $\mu^+$ and $\mu^-$ are produced with $\sigma_z=100$~$\mu$m and $\sigma_r=40$~$\mu$m.  Initial estimates indicate that such bunches of muons could be effectively accelerated given electric fields of $2$~GeV/m~\cite{ref:Alexey}, which can be achieved with LHC proton bunches.  
\item The FCC would be an excellent driver of plasma wakefields given the very high proton energy and the small bunch emittance~\cite{ref:FCC}.  Introducing long plasmas in the straight sections of the FCC would allow for the production of multi-TeV electron bunches, further greatly extending the physics capability of the FCC.  On the other hand, it may also be possible to accelerate electron and positron bunches to $50$~GeV or more in the straight sections without significant loss of protons, thus allowing for a high luminosity $ep$ and $e^+e^-$ programs at moderate additional cost.
\item Partially stripped ions can be cooled effectively in the LHC~\cite{ref:Krasny}, allowing for much higher luminosity $eA$ collisions for VHEeP.  If these ions can be phase rotated quickly in the LHC without significant loss, then short ion bunches could prove to be very effective drivers of plasma wakes.
\end{itemize}
There are surely many other possible applications of the PDPWA concept to be proposed and investigated.

\section{Summary}
Proton-driven plasma wakefield acceleration (PDPWA) is based on the transfer of energy from a proton bunch to a trailing bunch of particles, the `witness' particles, via plasma electrons.  The AWAKE scheme allows the use of existing proton accelerators for this purpose, and therefore extends the research opportunities made possible by the investments in the SPS and LHC.  Run~1 of AWAKE has ended in 2018 and all goals of the collaboration were met.  The seeded self-modulation process was observed to be robust and used to drive wakefields and accelerate electrons with high gradients.  AWAKE has now proposed a Run~2 to demonstrate that this scheme can be used for particle physics applications.  Run 2 is foreseen to take place between LS2 and LS3 of the LHC; i.e., from 2021-2024.  Given the continued successful development of the PDPWA scheme pursued by AWAKE, it will be possible to propose and realize particle physics projects based on this approach.  

First studies have been made on the use of electron bunches accelerated with an AWAKE-like scheme for particle physics projects.  As this document shows, the potential uses for particle physics projects of proton-driven plasma wakefield acceleration is clearly present and, given the success of AWAKE Run 2,  would become available in the midterm.  For this to become reality, strong support from the community is required now. 

 We note that the AWAKE scheme is also promising in conjunction with future accelerator concepts such as the FCC, the muon collider, and a gamma factory based on partially stripped ions.  The feasibility of for-purpose designed proton accelerators (with bunch lengths on the scale of the plasma wavelength) has not yet been investigated, but should be part of a long-term strategy.

\end{document}